\documentstyle[PASJadd, epsf]{PASJ95}

\draft
\begin{document}

\title{Infrared Properties of SiO Maser Sources in Late-Type Stars}

\author{Mikako {\sc Matsuura},$^{1,2}$
Issei {\sc Yamamura},$^{3,1}$
Hiroshi {\sc Murakami},$^1$
Takashi {\sc Onaka},$^2$ \\
Takafumi {\sc Ootsubo},$^2$
Takanao {\sc Tohya},$^{1,2}$
Yoshihiko {\sc Okamura},$^{1,4}$ \\
Minoru M.\ {\sc Freund},$^{1,5}$
and Masahiro {\sc Tanaka}$^1$
\\[12pt]
$^1${\it Institute of Space and Astronautical Science (ISAS),}\\
	{\it 3-1-1 Yoshinodai, Sagamihara, Kanagawa 229-8510}
\\
$^2${\it Department of Astronomy, University of Tokyo,}\\
	{\it 7-3-1 Hongo, Bunkyo, Tokyo 113-0033}\\
$^3${\it Astronomical Institute `Anton Pannekoek',
	University of Amsterdam,}\\
	{\it Kruislaan 403, 1098 SJ, Amsterdam, the Netherlands} \\
$^4${\it Department of Physics,	University of Tokyo,}\\ 
	{\it 7-3-1 Hongo, Bunkyo, Tokyo 113-0033} \\
$^5${\it Infrared Astrophysics Branch, Code 685,} \\
 {\it NASA Goddard Space Flight Center, Greenbelt, MD 20771, USA}
\\
{\it  E-mail (MM): mikako@astro.isas.ac.jp}} 

\abst{
Results of an SiO maser survey for the late-type stars
selected by the IRTS (Infrared Telescope in Space)
are presented.
We have detected
SiO $J=1-0,~v=1$ and/or $v=2$ lines in 27 stars out of 59 stars.
The maser intensity increases
with the depth of the H$_2$O absorption in the infrared spectra
and redness of the 2.2 and 12~$\mu$m color.
The column densities of the water vapor in the target stars are
estimated from the depth of the water absorption in the IRTS spectra.
We found that the SiO maser was detected mostly
in the stars with the column density of water vapor higher
than $3\times10^{19}$--$3\times10^{20}$~cm$^{-2}$.
We further estimate the density of hydrogen molecules 
in the outer atmosphere corresponding to these column densities, 
obtaining $10^{9}$--$10^{10}$~cm$^{-3}$ as a lower limit.
These values are roughly in agreement with the critical hydrogen density
predicted by models for the excitation of the SiO masers.
It is possible that the SiO masers are excited
in clumps with even higher than this density.
The present results provide useful information on the 
understanding of the
physical conditions of the outer atmospheres in late-type stars.
}

\kword{Stars: atmospheres --- Stars: late-type --- Radio sources: lines
--- Stars: long period variables --- Infrared: sources}

\maketitle
\thispagestyle{headings}

\section
{Introduction}
One of the characteristics of the red giant stars is
the possession of the atmosphere 
more extended than the hydrostatic case.
Several kinds of molecules are formed in the extended atmosphere.
Dust grains are also formed in the outermost region of the extended atmosphere,
and consequently, the mass loss is accelerated by the radiation pressure
on the dust grains.
The extended atmosphere is thought to be generated by large pulsations
of the stars.
Woitke et al. (1999) showed theoretically that 
pulsation affects the density structure of the molecules, especially
poly-atomic molecules.
However, poly-atomic molecules, such as
H$_2$O, CO$_2$, and SO$_2$,
are difficult to observe from the ground
because of the interference by the terrestrial atmosphere.
Recent observations based on satellite missions allowed investigations of the
properties of these molecules in the extended atmosphere
(Tsuji et al.\ 1997; Justtanont et al.\ 1998;
Ryde et al.\ 1999; Yamamura et al.\ 1999a, b).
Tsuji et al.\ (1997) found that a layer of molecules,
which the authors called a ``warm molecular envelope'',
is also present
in non-Mira variables,
i.e., irregular and semi-regular variables.
Pulsations of these stars are rather weak 
and thus it is under discussion at present
if the pulsation is still responsible
for the formation of the molecular layer even in non-Mira variables.
Hereafter, we denote the extended atmosphere or the warm molecular envelope
as an ``outer atmosphere'', which is
the region located above the photosphere,
but below the circumstellar envelope (see also Yamamura and de~Jong 2000).

Matsuura et al.\ (1999; hereafter Paper~I)
have studied the water vapor absorption 
band at 1.9~$\mu$m based on the
spectro-photometric
data obtained by the Near-Infrared Spectrometer (NIRS; Noda et al.\ 1996)
on board the Infrared Telescope in Space (IRTS; Murakami et al.\ 1996).
The NIRS observed water vapor absorption 
in a large sample of stars.
In the near infrared region, 
the water vapor is a dominant absorber in red-giants.
The water vapor absorption bands are thought to
arise mainly from the outer atmosphere of these stars.
The strength of the water vapor absorption increases with 
the 2.2 and 12~$\mu$m color, $C_{12/2.2}$.
This color 
is an indicator of the amount of dust in the inner part of
the circumstellar shell.
In Paper~I, we suggested that the depth of the water vapor absorption
represents the amount of matter in the outer atmosphere,
indicating the close relation between the density of the
outer atmosphere and the mass-loss rate.

SiO masers have been detected in many oxygen-rich
red-giants and red-supergiants.
SiO molecules are condensed into dust in the region with the temperature
around $\sim 1000$~K.
SiO masers
are thought to be excited in the region below the dust forming region
(Nyman and Olofsson 1986),
or in the outer atmosphere.
VLBI observations detect
SiO maser spots at 1--4 stellar radii from the central stars
(Miyoshi et al.\ 1994; Diamond et al.\ 1994).
The detection of these spots supports the idea proposed by
Langer and Watson (1984)
that the infalling wind makes inhomogeneous structures
and strong SiO maser emission comes from high-density clumps.
However, 
it is difficult to derive properties of the
outer atmosphere quantitatively
only from the observations of the SiO masers and
to understand the circumstances for the excitation of
SiO masers.
Two pumping mechanisms have been proposed so far
for the excitation of SiO masers;
radiative pumping 
(e.g. Kwan and Scoville 1974; Deguchi and Iguchi 1976; Bujarrabal 1994a, b)
and collisional pumping (e.g. Elitzur 1980; Doel et al.\ 1995).
At present it is still not settled which mechanism is dominant
in late-type stars.

 The near- and mid-infrared observations by the IRTS provide
useful information on the outer atmosphere.
It is worthwhile comparing the SiO maser properties with the
parameters derived from near- and mid-infrared
observations by the IRTS. 
Previous SiO maser surveys
(e.g. Allen et al.\ 1989; Izumiura et al.\ 1994)
were mostly based on the IRAS Point Source Catalog 
(IRAS-PSC; Joint IRAS Working Group 1988).
%
%
The SiO detected sources and non-detected sources are uniformly scattered
in the regions of oxygen-rich stars,
and are not clearly separated on the IRAS color-color diagram
(Haikala 1990).
The SiO maser is much more frequently detected in Mira variables
than in other variable types.
Only a few semi-regular and irregular variables exhibit 
the SiO maser activity.
The non-Mira SiO maser sources usually have a visual amplitude
larger than 2.5 mag (Alcolea et al.\ 1990), 
which is comparable to Mira variables.
Thus it is suggested that SiO maser excitation is related to 
the stellar pulsation
because stars with a larger visual amplitude are expected to
have stronger pulsation.
Alcolea et al.\ (1990) suggested that
fewer detections of the SiO maser in semi-regular variables
indicates less developed outer atmosphere in these stars
because the pulsation of semi-regulars is weaker than that of Miras.

 In this paper, we report the results of an SiO maser survey in a sample of
the IRTS point sources.
The IRTS provided a large number of stellar spectra in
near- and mid-infrared regions, unaffected by the terrestrial atmosphere.
The infrared colors and
the strength of the water absorption 
are compared with the SiO maser intensity.
We will discuss the conditions for the excitation of the SiO masers.

\section
{Observations}

 We made observations of the SiO masers
with the 45-m telescope at Nobeyama Radio Observatory from May 20
to 25, and from June 10 to 12, 1998. 
We simultaneously observed three SiO maser lines,
$J=1-0, v=1$ (43.1220~GHz),
$J=1-0, v=2$ (42.8205~GHz), and 
$J=2-1, v=1$ (86.2434~GHz). 
Two cooled SIS receivers (S40 for 43~GHz, and S100 for 86~GHz)
were used with acousto-optical
spectrometers (AOS's).
Each AOS covers 40~MHz with a resolution of 37~kHz.
The velocity coverage is
about $\pm$140~km~s$^{-1}$ for the $J=1-0$ lines, and
about $\pm$70~km~s$^{-1}$ for the $J=2-1$ line, respectively. 
The overall system temperature at 43~GHz
was typically 150--250~K with the worst case of 900K, depending
on the weather and the telescope elevation.
The $J=2-1$ line was observed
only when the conditions were appropriate, and the overall system temperature
ranged between 300 and 450~K. The half-power beam width of the telescope is 
40~arcsec at 43~GHz and 18~arcsec at 86~GHz, respectively.
Conversion factors from antenna temperature
($T_{\rm a}^{*}$ in K) to flux density (Jy) are
2.0~Jy~K$^{-1}$ at 43~GHz, and 2.6~Jy~K$^{-1}$ at 86~GHz.

 The target stars were selected from a preliminary list of the IRTS
point sources, whose
spectra were obtained by both the Mid-Infrared Spectrometer
(MIRS; Roellig et al.\ 1996),
and the NIRS.
The selection criteria are :
(1) The flux is approximately
larger than 10~Jy at 10~$\mu$m and larger than 1~Jy at 2.2~$\mu$m.
To get a wide coverage of different infrared colors and water indices,
we also include some stars with flux 
below 10~Jy at 10~$\mu$m;
(2) The spectrum shows the characteristics of
an oxygen-rich star (c.f. Yamamura et al.\ 1997).
Carbon-rich stars and S-type stars in the catalogues
(Stephenson 1984, 1989)
are excluded;
(3) The star has an identification in the IRAS-PSC.
In the present study,
we include some stars near the galactic plane,
which were not included in Paper~I.
The positions of the stars are taken from 
the IRAS-PSC,
the General Catalogues of Variable Stars
   (GCVS; Kholopov et al.\ 1988),
the Smithsonian Astrophysical Observatory Star Catalog
   (SAO; Smithsonian Institution 1966), 
and in a few cases from other catalogues.

\section{Results}

 The SiO $J=1-0, v=1$ and/or $v=2$ maser lines were
detected in 27 stars out of 59 target stars.
The infrared properties of detected sources and non-detected sources
are listed in tables~1 and 2. The results of SiO observations
are presented in tables~3 and 4.
The spectra of the SiO masers are shown in figure~1. 
The infrared colors $C_{2.2/1.7}$ and
$C_{12/2.2}$, and the water index $I_{\rm H_{2}O}$ in
table~1 and 2 are defined by:
 \begin{equation} 
    C_{2.2/1.7} = \log (F_{2.2} / F_{1.7}),
 \end{equation}
 \begin{equation}
    C_{12/2.2} = \log (F_{12} / F_{2.2}),
 \end{equation}
and
 \begin{equation} 
    I_{\rm H_{2}O} = \log (F_{\rm cont.} / F_{1.9}),
 \end{equation}
where $F_{2.2}$, $F_{1.7}$, $F_{1.9}$ are the IRTS/NIRS flux densities
in units of Jy at the 2.2, 1.7 and 1.9~$\mu$m channels, respectively.
$F_{\rm cont}$ is the continuum flux level at 1.9~$\mu$m,
which is calculated by linear
interpolation between $F_{1.7}$ and $F_{2.2}$. 
$F_{12}$ is the IRAS 12~$\mu$m flux.
$C_{2.2/1.7}$ is an indicator of spectral types 
for stars earlier than M6 (see Paper~I).
In the following discussion, we use the peak intensity of $J=1-0, v=2$
to represent the strength of the SiO masers of each star.
The integrated intensity
has almost a linear relation with the peak intensity.
The relation between the peak intensities of $J=1-0, v=2$
and $J=1-0, v=1$ lines is also mostly linear.

One irregular variable, TU~Lyr ($= \rm IRAS~18186+3143$),
shows a strong $J=2-1, v=1$ maser intensity,
while the $J=1-0, v=1$ and $v=2$ lines are
rather week.
This object has an unusually broad line-width
of about 20~km~s$^{-1}$
compared to the line-widths
less than 10~km~s$^{-1}$ for other stars.

%
%

%
%

%
%

%
%

%
%

%
%

%
%

\section{Detection Rate of the SiO Masers and the Infrared Properties}

\subsection{\it IRAS color-color diagram}

 The distribution of our sample on the IRAS color-color diagram
is shown in figure~2.
All the sampled stars have good IRAS flux quality at 12 and 25~$\mu$m.
Four stars with low quality at the 60~$\mu$m band are
not plotted in figure~2.
%
%
The division on the color-color diagram is taken from
van der Veen and Habing (1988). 
The sampled stars are mostly distributed in region II and IIIa, which
are the regions for oxygen-rich 
AGB stars with a low mass-loss rate of the order of $10^{-7}$
$\rm M_\solar$ yr$^{-1}$ (van der Veen and Habing 1988).
Several stars are located in region VII, which is the region for carbon stars.
However, the near-infrared spectra confirm their oxygen-rich nature.
`Detection' in figure~2
indicates either $J=1-0, v=1$ or $J=1-0, v=2$ lines are
detected in that star.
No clear distinction is seen 
between the detected sources and
the non-detected sources
on the IRAS color-color diagram.
This result is consistent with 
previous studies (e.g. Haikala 1990).

%
%

\subsection{Near-infrared color and water index}

 One of the purposes of this survey is to search for the SiO masers
in early M-type stars with water absorption reported in Paper~I. 
These stars are supposed to have developed an extended
atmosphere already among the early M-type stars observed by the NIRS,
and thus, SiO masers might be excited.
We observed two stars in this category visible from Nobeyama,
AK~Cap (M2) and IRAS~20073$-$1041 ($=$SAO~163310, M3).
However, SiO masers were not detected in both stars.

 Figure~3 shows the distribution of
the observed sources on the $I_{\rm H_2O}$ and $C_{2.2/1.7}$ diagram.
For reference, we plot several lines taken from figure~3 of Paper~I.
Three lines at the left bottom corner indicate 
the estimated zero level of the $I_{\rm H_2O}$ and $C_{2.2/1.7}$ relation
for stars without H$_2$O absorption,
and the $\pm 2\sigma$ deviations from that level. 
Stars above
$+2\sigma$ line clearly exhibit water absorption (see figure~4
in Paper~I).
%
%
$C_{2.2/1.7}$ is an indicator of spectral types for
stars earlier than M6.
Stars bluer than $C_{2.2/1.7}<-0.085$ (vertical line)
are expected to have spectral types earlier than M6.
One M6 star, IRAS~19461+0334 (WX~Aql), which was not included in Paper~I,
is actually located leftward of this vertical line, 
but this will not affect our conclusions in this paper.
In figure~3, a clear separation is seen
between the SiO detected sources and the SiO non-detected sources.
SiO masers are detected in stars with strong H$_2$O absorption
($I_{\rm H_2O}$ is approximately larger than 0.1).
In addition, stars in the region of
$I_{\rm H_2O}$ larger than $\sim0.1$ are mostly
Mira variables when their variable types are known.
SiO masers are seldom detected
in non-Mira variables (see Habing 1996).
The present study confirms the same trend.
The numbers of the SiO maser detected sources and non-detected sources
for different variable types are summarized in table~5.
SiO masers are detected with a high probability in Mira variables,
while they are hardly detected in SRs and Lbs.

%
%

%
%

\section{The Relation between SiO Maser Intensity and Infrared Parameters}

\subsection{Near- and mid-infrared color}
 The infrared color $K-[12]$ is considered as
an indicator of mass-loss rate
for Mira variables
(Whitelock et al.\ 1994; Le~Sidaner and Le~Bertre 1996).
We use $C_{12/2.2}$ instead of $K-[12]$,
because the $F_{2.2}$ band represents the continuum level in $K$.
The relation between the SiO maser intensity and the $C_{12/2.2}$
is shown in figure~4.
We plot the SiO maser intensity divided by $F_{2.2}$.
We implicitly assume that 
the intrinsic 2.2~$\mu$m band luminosities of the sample stars are similar,
and that the extinction by circumstellar dust is small in the 2.2~$\mu$m band,
except for a few extremely red stars.
In figure~4, the SiO maser intensity seems to correlate 
with $C_{12/2.2}$ in the color range $C_{12/2.2} < 0.5$. 
Thus SiO maser intensity increases with mass-loss rate.
The color $C_{12/2.2} = +0.5$ corresponds to a dust mass-loss 
rate of approximately ${\rm 7\times10^{-9}~M_{\solar}~yr^{-1}}$.
Here we use the equations in 
Le~Sidaner and Le~Bertre (1993, 1996)
and the zero magnitude flux of $F_{2.2}$
is equal to 625~Jy (Cohen 1997).
We assume the inner radius of the dust shell of
$2.5 \times 10^{14}$~cm
and the expanding velocity of 10~km~s$^{-1}$.

 Nyman and Olofsson (1986) suggested that
the SiO maser intensity does not correlate with mass-loss rate. They use
a mass-loss rate calculated from CO $J=1-0$ thermal emission. 
The CO emission is usually dominated by the molecules in
the outer region of the circumstellar envelope
and it represents
a mass-loss rate averaged over the past thousands years.
On the other hand, $K-[12]$ or $C_{12/2.2}$ represents emission from
hot dust in the inner envelope and thus these colors indicate a recent
mass-loss rate of the star.
Since SiO masers are thought to be
excited just below the dust forming region,
it is likely
that the SiO maser intensity correlates better with $C_{12/2.2}$
than the CO emission.

%
%

\subsection{Water index}

 The SiO maser intensity is shown as a function of $I_{\rm H_2O}$ in figure~5.
In general, the SiO maser intensity among the detected stars
increases with $I_{\rm H_2O}$,
but the scatter is large.
The large scatter might come from
the time variation of the water absorption and the SiO maser intensity.
In Mira variables, the water absorption depth changes from phase to phase
(Hyland 1974), and the column density of water changes by
a factor of 10
(Hinkle and Barnes 1979).
The SiO maser intensity also varies by up to a factor of 10 
(Nyman and Olofsson 1986; Alcolea et al.\ 1999).
The variabilities in both quantities might obscure the relation
between the water index and the SiO maser intensity.

%
%

\section{Discussion}

 Figures~3 and 5 indicate that
the sources showing the SiO maser lines have 
$I_{\rm H_2O}$ larger than $\sim0.1$
and they are mostly Mira variables.
Since the water index, $I_{\rm H_2O}$, roughly indicates
the column density of water molecules in the outer atmosphere,
it is suggested
that water column density is systematically different between the SiO maser
detected stars and non-detected stars,
and also between Miras and non-Miras.
The SiO maser is known to be detected quite often
in Mira variables, while it is hardly detected in non-Miras
(see Habing 1996).
Alcolea et al.\ (1990) interpreted it in terms of the idea that
the outer atmosphere is more `developed' in Miras than in non-Miras.
Miras have stronger pulsation than non-Miras.
Stronger pulsation will lift up more matter
from the photosphere into the outer atmosphere,
leading to the higher density and thus higher column density of the molecules.
Our results of the relation with $I_{\rm H_2O}$
indicate that the high detection rate of the SiO maser in Miras
is related to the high column density in the outer atmosphere.

The column densities of water molecules have been measured
for several stars.
In table~6, we summarize the results of near-infrared observations of
the H$_2$O bands reported in Paper~I and other literatures.
In Paper~I we estimated the column density ($N$)
and the excitation temperature ($T_{\rm ex}$)
of the water molecules in two early M-type stars, AK~Cap and V~Hor.
The water spectra of these stars were fitted by
synthesized spectra using a plane-parallel model.
We apply the same method for SAO~163310 (M3)
and the results are indicated in table~6.
The column densities of these stars
are as uncertain as one order of the magnitude 
due to the low spectral resolution of the IRTS/NIRS
($\lambda / \Delta \lambda \approx$ 15).
These three early M-type stars show relatively strong H$_2$O absorption bands
among the stars with similar spectral types (Paper~I).
Therefore, the resultant column densities are
rather high compared to the other stars of similar types.
Hinkle and Barnes (1979) and Yamamura et al.\ (1999b) 
interpreted the observations of the water bands
in the Mira variables in terms of two molecular layers.
We list the parameters of both layers in table~6.
Hinkle and Barnes (1979) analyzed high-resolution spectra
($\lambda / \Delta \lambda =$ a few tens of thousands) 
of the Mira variable, R~Leo, at $K$ and $H$ bands, 
and found two radial velocity components of water lines.
One component has an excitation temperature of 1750~K
and the other component about 1150~K.
Yamamura et al. (1999b) fitted the water bands
taken by the ISO/SWS (Short-Wavelength Spectrometer)
in $o$~Cet and Z~Cas using their `slab' model consisting of two water vapor layers.
They stated that the layer with $T_{\rm ex}=2000$~K
should be as large as 2~$R_{\star}$ in order to reproduce
the emission band seen at 3.5--4.0~$\mu$m in $o$~Cet.
The cool layer is responsible for the absorption feature
between 2.5 and 3.5 $\mu$m.
In the case of Z~Cas, the hot layer is located around 1~$R_{\star}$
and both hot and cool layers contribute to absorption.
Hinkle and Barnes (1979) mention that the 1750~K layer
is located near the photosphere.
This layer may be identical to the 2000~K layer at 1--2~$R_{\star}$
in Yamamura et al.\ (1999b).
The cool layer with $T_{\rm ex}=1200$--1400~K 
of Miras in Yamamura et al.\ (1999b) is as large as 2~$R_{\star}$ or larger.
Tsuji et al.\ (1997) also showed that the layer with
an excitation temperature
of about 1000~K is located at $\sim 2~R_{\star}$
in irregular and semi-regular variables.
We note that 
according to the `slab' model,
when the water layer with
$T_{\rm ex}=1500$~K is extended to $3~R_{\star}$,
or the layer with $T_{\rm ex}=1000$~K to $5~R_{\star}$,
the water spectra should be observed in emission around 2.7~$\mu$m,
which is not the case for the present sample.
We suppose that 
the layers with temperatures between 1000 and 1500~K
in the stars in table~6
lie around 2~$R_{\star}$.

 Table~6 shows that the stars with the SiO maser have the
water column density higher than 
$3\times10^{19}$--$3\times10^{20}$~cm$^{-2}$,
and non-detected stars have column density 
lower than $5\times10^{19}$~cm$^{-2}$.
The column densities in table~6 
vary by more than one order of magnitude among the stars.
It is difficult to extend the water layer with the excitation temperature
above 1000~K in such a large scale as discussed above.
Thus, the large difference in the water column density should be
attributed to the difference in the water 
density in the stars.
The density of the hydrogen molecules in the outer atmosphere
can be estimated by dividing the water column densities
by H$_2$O abundance and the thickness of the layer.
Analysis based on the spherical model should be
required to make more accurate estimate,
but in this paper we use the numbers derived from 
a slab model for a rough estimate.
The density is probably inhomogeneous in the outer atmosphere
and the SiO maser might be excited in the clumps with high-density
in the outer atmosphere as suggested by Langer and Watson (1984).
We try to estimate a lower limit of the density.
The H$_2$O abundance ratio calculated with
thermal equilibrium 
is in the order of $10^{-4}$ (Tsuji 1964).
It depends on the abundance in the atmosphere.
As representative value, we adopt relatively large H$_2$O abundance
of Barlow et al.\ (1996), H$_2$O/H$_2$ $=8\times 10^{-4}$.
We assume that the size of the H$_2$O layer
is 1 $R_{\star}$ in line of sight,
based on the estimated 
location of the water layer of 1000--1500~K at $\sim 2~R_{\star}$.
The stellar radius $R_{\star}$ is 
between $3\times 10^{13}$ and $5\times10^{13}$~cm
in Mira variables (Tuthill et al.\ 1993; Tuthill et al.\ 1994),
and is estimated as
$3\times10^{13}$~cm for a semi-regular variable
using Tuthill, Haniff and Baldwin (1999) and
Hipparcos and Tycho Catalogues (1997).
We adopt $5\times10^{13}$~cm for the stellar radius.
With these values
we obtain $10^{9}$--$10^{10}$~cm$^{-3}$ for a lower limit of the
molecular hydrogen density corresponding to
the water column density of $3\times10^{19}$--$3\times10^{20}$~cm$^{-2}$.

The estimated number of $n(\rm{H}_2)=10^{9}$--$10^{10}$~cm$^{-3}$
is roughly in agreement with
the hydrogen density
where recent theoretical works 
predict that the maser intensity increases drastically
(Bujarrabal 1994a; Doel et al.\ 1995).
Bujarrabal (1994a) showed
that the SiO maser intensity
increases by four orders of magnitudes at the hydrogen densities
of $n({\rm H_2}) \sim 10^{8}$--$10^{9}$ cm$^{-3}$.
Doel et al.\ (1995) also showed that the
maser gain coefficient (gain per unit amplification path length)
increases steeply at $n({\rm H_2}) \sim 5\times10^{9}$ cm$^{-3}$.
Therefore,
the probability of the SiO maser detection
may be prescribed by the critical density
of about $10^{9}$--$10^{10}$~cm$^{-3}$.
However, the critical density at the `spots' where the SiO maser
is excited may be even higher if the maser
is excited in high-density clumps
(Langer and Watson 1984).

 Figure~4 shows the correlation between $C_{12/2.2}$
and the SiO maser intensity, which implies
the relation between the density
and the excitation of the SiO masers.
$C_{12/2.2}$ is a measure of the dust mass-loss rate
in the innermost region of the circumstellar envelope.
The mass-loss rate in this region is expected to increase with
the density in the outer atmosphere.
Therefore, the correlation between $C_{12/2.2}$ and SiO maser intensity
also implies that the maser excitation is related to
the density in the outer atmosphere.

 Apart from the density, there might be other possibilities
which affect the amplification of the SiO masers
and lead to the dependence of detection rate on the variable types.
For example, Hinkle, Lebzelter and Scharlach (1997) show 
that the amplitude of velocity variation 
depends on the amplitude of visual magnitudes.
The velocity structure
might vary depending on the strength of pulsation,
which could result in the coherent path length for
maser amplification.
Further observational studies in the infrared region
with high resolution observations are required
to investigate the effects of velocity structure on the maser excitation.

%
%

\section{Summary}

 The IRTS provided a large number of stellar spectra
without the interference from the terrestrial atmosphere.
These data enable us a systematic study
of the relation between the SiO maser excitation
and the conditions of the outer atmosphere.
The IRTS measured the water index $I_{\rm H_2O}$
in red giants,
which is an indicator of the water column density
in the outer atmosphere.
In 59 stars selected from the objects observed by the IRTS,
the SiO maser lines were detected in 27 stars.
Stars with deep water absorption 
($I_{\rm H_2O}$ larger than $\sim0.1$) mostly
show the SiO maser lines, and 
they are mostly Mira variables among those with known spectral type.
The SiO maser intensity is found to increase 
with the color $C_{12/2.2}$ and $I_{\rm H_2O}$.

Stars with $I_{\rm H_2O}$ larger than $\sim 0.1$ mostly
show SiO masers,
and the stars with SiO maser activity
have the water column density 
higher than $3\times10^{19}$--$3\times10^{20}$~cm$^{-2}$,
while non-detected stars show smaller than $5\times10^{19}$~cm$^{-2}$.
A lower limit of the molecular hydrogen 
density corresponding to the water column density
is estimated as $10^{9}$--$10^{10}$~cm$^{-3}$.
This number is roughly comparable with the critical
gas density predicted by the models,
where the SiO masers are excited
(Bujarrabal 1994a; Doel et al.\ 1995).
If the SiO masers are excited
in high-density clumps (Langer and Watson 1984),
the critical density would be even higher than
the values we derived here.

 The color $C_{12/2.2}$ well-correlates with the SiO maser intensities.
$C_{12/2.2}$ is probably influenced by the density of the outer atmosphere.
The relation $C_{12/2.2}$ with the SiO masers also suggests that
the density of the outer atmosphere is one of the key parameters
in the SiO maser excitation.

\par
\vspace{1pc}\par
We would like to thank the staff of Nobeyama Radio Observatory for their
supports.
We also appreciate 
Drs. S. Deguchi, H. Izumiura, and T. Fujii
for their suggestions for the observations.
M.M. thanks the Research Fellowships of the 
Japan Society for the Promotion of Science for the Young Scientists.
I.Y. acknowledges the financial support from a NWO PIONIER grant.

\section*{References}

\re
  Alcolea J., Bujarrabal V., 
  G$\rm \acute{o}$mez-Gonz$\rm \acute{a}$lez J.\ 1990,
  A\&A 231, 431

\re
  Alcolea J, Pardo J.R., Bujarrabal V., Bachiller R., Barcia A.,
  Colomer F., Gallego J.D., G$\rm \acute{o}$mez-Gonz$\rm \acute{a}$lez J.
  et al.\ 1999 A\&AS 139, 461

\re
  Allen D.A., Hall P.J., Norris R.P.,
  Troup E.R., Wark R.M., Wright A.E.\
  1989, MNRAS 236, 363


\re
  Barlow M.J., Nguyen-Q-Rieu, Truong-Bach, 
  Cernicharo J., Gonzalez-Alfonso E., Liu X.W.,
  Cox P., Sylvester R.J.
  et al.\ 
  1996, A\&A 315, L241


\re
  Buhl D., Snyder L.E., Lovas F.J., Johnson D.R.\ 1974, ApJ 192, L97


\re
  Bujarrabal V.\ 1994a, A\&A 285, 953

\re
  Bujarrabal V.\ 1994b, A\&A 285, 971


\re
  Cohen M.\ 1997
  in Proceedings for `Diffuse Infrared Radiation and the IRTS', 
  ASP Conference Series, vol.124, ed., 
  H. Okuda, T. Matsumoto, T.L. Roellig, p.~61

\re
  Deguchi S., Iguchi T.\ 1976, PASJ 28, 307

\re
  Diamond P.J., Kemball A.J., Junor W., Zensus A.,
  Benson J., Dhawan V.\
  1994, ApJ 430, L61


\re
  Doel R.C., Gray M.D., Humphreys E.M.L., Braithwaite M.F.,
  Field D.\ 1995, A\&A 302, 797

\re
  Elitzur M.\ 1980, ApJ 240, 553


\re
  Herpin F., Baudry A., Alcolea J., Cernicharo J.\
  1998, A\&A 334, 1037

\re
  Habing H.J.\ 1996, A\&ARv 7, 97

\re
  Heske E.\ 1989, A\&AS 208, 77

\re
  Haikala L.K.\ 1990 A\&AS 85, 875



\re
  Hinkle K.H., Barnes T.G.\ 1979, ApJ 227, 923

\re
  Hinkle K.H., Lebzelter T., Scharlach W.W.G.\ 1997
  AJ 114, 2686

\re
  The Hipparcos and Tycho Catalogue 1997, ESA




\re 
  Hyland A.R., 1974, Highlights of Astronomy, vol. 3, 307, IAU

\re
  Justtanont K., Feuchtgruber H., de Jong T., Cami J.,
  Waters L.B.F.M., Yamamura I., Onaka T.\ 1998, A\&A 330, L17

\re
  IRAS Point Source Catalog 1988, Joint IRAS Working Group, Washington DC.
  (IRAS-PSC)

\re
  Izumiura H., Deguchi S., Hashimoto O., Nakada Y., Onaka T., Ono T.
  Ukita N., Yamamura I.\
  1994, ApJ 437, 419

\re
  Kholopov P.N., Samus N.N., Frolov M.S., Goranskij V.P., 
  Gorynya N.A., Kireeva N.N., Kukarkina N.P., Kurochkin N.E.
  et al.\ 1988, 
 `General Catalogue of
 Variable Stars, 4th Ed.' (Nauka Publishing House, Moscow) (GCVS)

\re
  Kwan J., Scoville N.\ 1974, ApJ 194, L97

\re
  Langer S.H. \& Watson W.D. 1984, 284, 751

\re
  Le~Sidaner P., Le~Bertre T.\ 1993, A\&A 278, 167

\re
  Le~Sidaner P., Le~Bertre T.\ 1996, A\&A 314, 896

\re 
  Matsuura M., Yamamura I., Murakami H., Freund M.M., Tanaka M.\
  1999, A\&A 348, 579 (Paper~I)

\re
  Miyoshi M., Matsumoto K., Kameno S., Takaba H., Iwata T.\
  1994, Nature 371, 395

\re
  Murakami H., Freund M.M., Ganga K., Guo H.F., Hirao T., Hiromoto N.,
  Kawada M., Lange E.A. et al.\ 1996, PASJ 48, L41

\re
  Noda M., Matsumoto T., Murakami H., Kawada M., Tanaka M.,
  Matsuura S., Guo H.F.\ 1996, SPIE 2817, 248

\re
  Nyman L.$\rm{\AA}$., Olofsson H.\ 1986, A\&A 158, 67



\re
  Roellig T., Mochizuki K., Onaka T., Tanabe T.,
  Yamamura I., Yuen L.M.\ 1996, SPIE 2817, 258

\re
  Ryde N., Eriksson K., Gustafson B.\ 1999, A\&A 341, 579



\re
  Smithsonian Astrophysical Observatory Star catalog 1966, 
  Smithsonian Institution, Washington DC. (SAO)

\re
  Spencer J.H., Winnberg A., Olnon F.M., Schwartz P.R.,
  Matthews H.E., Downes D.\
  1981 AJ, 86, 329

\re
  Stephenson C.B.\ 1984, `General Catalog of S Stars',
2nd ed., Publication of 
Warner \& Swasey Observatory

\re
  Stephenson C.B.\ 1989,
`General Catalog of Cool Galactic Carbon Stars', 2nd ed., 
Publication of Warner \& Swasey Observatory

\re
  Tuthill P.G., Haniff C.A., Baldwin J.E.\
  1993,
  in Proceedings for `Very High Angular Resolution Imaging', 
  IAU Symp.158 ed., J.G. Robertson, W.J. Tango, p.~395

\re
  Tuthill P.G., Haniff C.A., Baldwin J.E., Feast M.W.\
  1994, MNRAS 266, 745

\re
  Tuthill P.G., Haniff C.A., Baldwin J.E.\
  1999, MNRAS 306, 353


\re
  Tsuji T. 1964 Ann. Tokyo Astr. Obs., 9, 1


\re
  Tsuji T., Ohnaka K., Aoki W., Yamamura I.\ 1997, A\&A 320, L1

\re
  Van der Veen W.E.C.J, Habing H.J.\ 1988, A\&A 194, 125

\re
  Whitelock P., Menzies J., Feast M., Marang F., Carter B.,
  Roberts G., Catchpole R., Chapman J.\
  1994, MNRAS 267, 711

\re
  Woitke P., Helling Ch., Winters J.M., Jeong K.S.\
  1999, A\&A 348, L17

\re
  Yamamura I. and the IRTS Team 1997
  in Proceedings for `Diffuse Infrared Radiation and the IRTS', 
  ASP Conference Series, vol.124, ed., 
  H. Okuda, T. Matsumoto,  T.L. Roellig, p.~72

\re
  Yamamura I., de Jong T., Onaka T., Cami J., Waters L.B.F.M.\
  1999a, A\&A 341, L9

\re
  Yamamura I., de Jong T., Cami J.\ 1999b, A\&A 348, L55

\re 
  Yamamura I., de Jong T.\ 2000, ESA SP-456, in press

\label{last}

%
%

\clearpage
\centerline{
}
\bigskip
\begin{fv}{1.a}{25cm}
{
SiO maser spectra of detected sources are shown. 
From top to bottom, spectra indicate $J=1-0, v=1$;
$J=1-0, v=2$; and $J=2-1, v=1$ lines.
When $J=2-1, v=1$ line was not observed, only 
two $J=1-0$ spectra are presented.
IRAS~$18186+3143$ was observed three times,
and every spectra are plotted.
}
\end{fv}

%
%

\clearpage
\centerline{}
\bigskip
\begin{fv}{1.b}{25cm}
{Same as figure~1.a.}
\end{fv}

%
%

\clearpage
\centerline{
}
\bigskip
\begin{fv}{2}{10cm}
{
Observed stars of our SiO maser survey are plotted on the
IRAS color-color diagram.
The division of the regions comes from
van der Veen \& Habing (1988).
The SiO detected stars are not well separated from the non-detected stars.
}
\end{fv}

%
%

\clearpage
\centerline{
}
\bigskip
\begin{fv}{3}{7cm}
{
The water index $I_{\rm H_2O}$ is plotted as a function of $C_{2.2/1.7}$.
The lines are from Paper~I (see text).
Symbols indicate SiO $J=1-0$ lines detection/non-detection, 
and variable types.
The triangles on the upper side indicate the average color
at each spectral type, derived from the sample in Paper~I.
}
\end{fv}

%
%

\clearpage
\centerline{}
\begin{fv}{4}{10cm}
{
SiO $J=1-0, v=2$ maser intensity is plotted against $C_{12/2.2}$.
For the non-detected stars, 3$\sigma$ (1$\sigma$ is the $rms$ level)
is used for maser intensity, and this indicates the upper limit.
To make the parameter independent from the distance,
SiO maser intensity is divided by $F_{2.2}$.
$C_{12/2.2}$ is an indicator of the mass-loss rate.
There is a correlation between these indices
in low mass-loss rate stars ($C_{12/2.2}<0.5$,
approximate dust mass loss rate of $7\times10^{-9}~{\rm M_{\solar}~yr^{-1}}$).
}
\end{fv}

%
%

\clearpage
\centerline{
}
\begin{fv}{5}{10cm}
{
The relation between
water index $I_{\rm H_2O}$ and SiO maser intensity is indicated.
For non-detected sources, we use the 3 $\sigma$ levels 
for $T_{\rm a}^*$.
Only a weak relation is seen between $I_{\rm H_2O}$ and SiO maser intensity.
}
\end{fv}


%
%

\begin{table}[t]
\scriptsize
\begin{center}
Table~1. \hspace{4pt} List of SiO $J=1-0, v=$1 and/or $v=2$ masers 
detected sources.
\end{center}
\vspace{6pt}
\begin{tabular*}{\columnwidth}{@{\hspace{\tabcolsep}
\extracolsep{\fill}}l ll rr lc c}
\hline\hline\\[-6pt]
IRAS name  & R.A. (B.1950)& Dec (B.1950) & ${I_{\rm H_2O}}$ & \it{C$_{2.2/1.7}$} & GCVS name & Type & Comments \\[4pt]\hline\\[-6pt]

02255$+$6903 \dotfill & $02^{\rm h}25^{\rm m}33^{\rm s}\hspace{-5pt}.\hspace{2pt}8$ & $+69^{\rm d}03^{\rm m}36^{\rm s}\hspace{-5pt}.\hspace{2pt}0$  & 0.074 &$ 0.015$&  &  &\\ 
04081$+$5832 \dotfill & $04^{\rm h}08^{\rm m}06^{\rm s}\hspace{-5pt}.\hspace{2pt}8$ & $+58^{\rm d}32^{\rm m}22^{\rm s}\hspace{-5pt}.\hspace{2pt}0$  & 0.171 &  0.015 &  &  &\\  
06266$-$1148 \dotfill & $06^{\rm h}26^{\rm m}36^{\rm s}\hspace{-5pt}.\hspace{2pt}1$ & $-11^{\rm d}48^{\rm m}58^{\rm s}\hspace{-5pt}.\hspace{2pt}0$  & 0.286 &  0.084 &  &  &\\  
07019$-$1631 \dotfill & $07^{\rm h}01^{\rm m}58^{\rm s}\hspace{-5pt}.\hspace{2pt}0$ & $-16^{\rm d}31^{\rm m}02^{\rm s}\hspace{-5pt}.\hspace{2pt}0$  & 0.328 &  0.007 &  &  &\\  
07232$-$0544 \dotfill & $07^{\rm h}23^{\rm m}13^{\rm s}\hspace{-5pt}.\hspace{2pt}0$ & $-05^{\rm d}44^{\rm m}59^{\rm s}\hspace{-5pt}.\hspace{2pt}0$  & 0.234 &$-0.047$& TT MON & M &1)\\  
07268$-$0410 \dotfill & $07^{\rm h}26^{\rm m}52^{\rm s}\hspace{-5pt}.\hspace{2pt}0$ & $-04^{\rm d}10^{\rm m}30^{\rm s}\hspace{-5pt}.\hspace{2pt}0$  & 0.022 &$-0.035$& RX MON & M &\\   
08186$+$1409 \dotfill & $08^{\rm h}18^{\rm m}36^{\rm s}\hspace{-5pt}.\hspace{2pt}6$ & $+14^{\rm d}09^{\rm m}49^{\rm s}\hspace{-5pt}.\hspace{2pt}0$  & 0.172 &$-0.035$& SZ CNC & M &\\  
12341$+$5945 \dotfill & $12^{\rm h}34^{\rm m}07^{\rm s}\hspace{-5pt}.\hspace{2pt}0$ & $+59^{\rm d}45^{\rm m}42^{\rm s}\hspace{-5pt}.\hspace{2pt}9$  & 0.192 &$-0.060$& T UMA & M &\\  
18076$+$3445 \dotfill & $18^{\rm h}07^{\rm m}37^{\rm s}\hspace{-5pt}.\hspace{2pt}0$ & $+34^{\rm d}45^{\rm m}40^{\rm s}\hspace{-5pt}.\hspace{2pt}0$  & 0.301 &  0.288 &  &  &\\ 
18156$+$0655 \dotfill & $18^{\rm h}15^{\rm m}40^{\rm s}\hspace{-5pt}.\hspace{2pt}8$ & $+06^{\rm d}55^{\rm m}01^{\rm s}\hspace{-5pt}.\hspace{2pt}0$  & 0.228 &  0.002 & BC OPH & M & \\ 
18186$+$3143 \dotfill & $18^{\rm h}18^{\rm m}37^{\rm s}\hspace{-5pt}.\hspace{2pt}6$ & $+31^{\rm d}43^{\rm m}54^{\rm s}\hspace{-5pt}.\hspace{2pt}0$  & 0.101 &$-0.055$& TU LYR & Lb & \\  
18222$+$3933 \dotfill & $18^{\rm h}22^{\rm m}18^{\rm s}\hspace{-5pt}.\hspace{2pt}0$ & $+39^{\rm d}33^{\rm m}24^{\rm s}\hspace{-5pt}.\hspace{2pt}0$  & 0.108 &$-0.027$& TW LYR & M &1)\\  
18347$+$2600 \dotfill & $18^{\rm h}34^{\rm m}45^{\rm s}\hspace{-5pt}.\hspace{2pt}0$ & $+26^{\rm d}00^{\rm m}24^{\rm s}\hspace{-5pt}.\hspace{2pt}0$  & 0.081 &$-0.039$& RZ HER & M &\\ 
18394$+$2845$^\diamondsuit$ \dotfill & $18^{\rm h}39^{\rm m}29^{\rm s}\hspace{-5pt}.\hspace{2pt}9$ & $+28^{\rm d}45^{\rm m}56^{\rm s}\hspace{-5pt}.\hspace{2pt}4$  & 0.050 &$-0.070$& SY LYR & SRb &\\ 
18561$+$1642 \dotfill & $18^{\rm h}56^{\rm m}10^{\rm s}\hspace{-5pt}.\hspace{2pt}0$ & $+16^{\rm d}42^{\rm m}48^{\rm s}\hspace{-5pt}.\hspace{2pt}0$  & 0.174 &$-0.028$& EU AQL & M &2)\\ 
19090$+$1746 \dotfill & $19^{\rm h}09^{\rm m}03^{\rm s}\hspace{-5pt}.\hspace{2pt}6$ & $+17^{\rm d}46^{\rm m}48^{\rm s}\hspace{-5pt}.\hspace{2pt}0$  & 0.108 &  0.024 & KK SGE & M &2)\\ 
19158$+$1955 \dotfill & $19^{\rm h}15^{\rm m}48^{\rm s}\hspace{-5pt}.\hspace{2pt}9$ & $+19^{\rm d}55^{\rm m}54^{\rm s}\hspace{-5pt}.\hspace{2pt}0$  & 0.103 &$-0.001$& NO SGE & M &2)\\ 
19172$+$1706 \dotfill & $19^{\rm h}17^{\rm m}18^{\rm s}\hspace{-5pt}.\hspace{2pt}0$ & $+17^{\rm d}06^{\rm m}48^{\rm s}\hspace{-5pt}.\hspace{2pt}0$  & 0.292 &  0.001 & W SGE & M &2)\\ 
19285$+$4853 \dotfill & $19^{\rm h}28^{\rm m}33^{\rm s}\hspace{-5pt}.\hspace{2pt}9$ & $+48^{\rm d}53^{\rm m}45^{\rm s}\hspace{-5pt}.\hspace{2pt}0$  & 0.101 &$-0.046$&  &  &\\  
19308$+$0609$^\diamondsuit$ \dotfill & $19^{\rm h}30^{\rm m}55^{\rm s}\hspace{-5pt}.\hspace{2pt}0$ & $+06^{\rm d}09^{\rm m}30^{\rm s}\hspace{-5pt}.\hspace{2pt}0$  & 0.059 &$-0.079$& V 621AQL & SRb &\\ 
19412$+$0337 \dotfill & $19^{\rm h}41^{\rm m}15^{\rm s}\hspace{-5pt}.\hspace{2pt}3$ & $+03^{\rm d}37^{\rm m}15^{\rm s}\hspace{-5pt}.\hspace{2pt}0$  & 0.402 &  0.236 & $^*$ &  &\\ 
19550$-$0201 \dotfill & $19^{\rm h}55^{\rm m}01^{\rm s}\hspace{-5pt}.\hspace{2pt}0$ & $-02^{\rm d}01^{\rm m}12^{\rm s}\hspace{-5pt}.\hspace{2pt}0$  & 0.357 &  0.048 & RR AQL & M &\\ 
20077$-$0625 \dotfill & $20^{\rm h}07^{\rm m}46^{\rm s}\hspace{-5pt}.\hspace{2pt}0$ & $-06^{\rm d}24^{\rm m}42^{\rm s}\hspace{-5pt}.\hspace{2pt}0$  & 0.336 &  0.392 & V1300AQL$^\dagger$ & M: & 1)\\ 
20215$+$6243 \dotfill & $20^{\rm h}21^{\rm m}32^{\rm s}\hspace{-5pt}.\hspace{2pt}3$ & $+62^{\rm d}43^{\rm m}21^{\rm s}\hspace{-5pt}.\hspace{2pt}0$  & 0.097 &$-0.004$&  &  & \\ 
20234$-$1357 \dotfill & $20^{\rm h}23^{\rm m}26^{\rm s}\hspace{-5pt}.\hspace{2pt}0$ & $-13^{\rm d}57^{\rm m}51^{\rm s}\hspace{-5pt}.\hspace{2pt}0$  & 0.236 &  0.061 &  &  & \\ 
20296$-$2151 \dotfill & $20^{\rm h}29^{\rm m}38^{\rm s}\hspace{-5pt}.\hspace{2pt}7$ & $-21^{\rm d}51^{\rm m}40^{\rm s}\hspace{-5pt}.\hspace{2pt}0$  & 0.332 &  0.016 & RU CAP & M &\\ 
20305$+$6246 \dotfill & $20^{\rm h}30^{\rm m}35^{\rm s}\hspace{-5pt}.\hspace{2pt}4$ & $+62^{\rm d}46^{\rm m}28^{\rm s}\hspace{-5pt}.\hspace{2pt}0$  & 0.078 &  0.014 & BF CEP & M &\\ 

\hline
\end{tabular*}
$\diamondsuit$ indicates tentative detection.
`Type' is variable type according to GCVS.
`M' : Mira variable.
`SR' : semi-regular variable,
`SRb' : a sub-class of `SR' and indicate 
semi-regular variable with `poor defined periodicity'.
`Lb' : irregular variable.
Comments show the quality of NIRS spectra.
1) Confusion with other source occurred, i.e 
 more than 2 source entered in the one 
 field of view when the star was observed, because of the NIRS's
 large aperture.
 Brightness of the fainter sources are about 20\% of the target star at 2.2~$\mu$m.
2) The flux quality is low,
 because of the different satellite operation near the galactic plane.
 About 10 -- 20 \% flux uncertainty remains.

\vspace{6pt}\par\noindent
$*$ IRC$+00450$
$\dagger$ IRC$-10529$

\end{table}

%
%

\begin{table}[t]
\small
\scriptsize

\begin{center}
 Table~2. \hspace{4pt} List of SiO non-detected sources.
\end{center}

\vspace{6pt}
\begin{tabular*}{\columnwidth}{@{\hspace{\tabcolsep}
\extracolsep{\fill}}l ll rr lc c}
\hline\hline\\[-6pt]
IRAS name  & R.A. (B.1950)& Dec (B.1950) & ${I_{\rm H_2O}}$ & \it{C$_{2.2/1.7}$} & GCVS name & Type & Comments\\[4pt]\hline\\[-6pt]

01010$+$7434 \dotfill & $01^{\rm h}01^{\rm m}03^{\rm s}\hspace{-5pt}.\hspace{2pt}8$ & $+74^{\rm d}34^{\rm m}00^{\rm s}\hspace{-5pt}.\hspace{2pt}1$ &  0.013 &$-0.063$&          &     &\\
01584$+$7103 \dotfill & $01^{\rm h}58^{\rm m}26^{\rm s}\hspace{-5pt}.\hspace{2pt}8$ & $+71^{\rm d}03^{\rm m}24^{\rm s}\hspace{-5pt}.\hspace{2pt}0$ &$-0.005$&$-0.105$& &  &\\
03478$+$6349 \dotfill & $03^{\rm h}47^{\rm m}53^{\rm s}\hspace{-5pt}.\hspace{2pt}3$ & $+63^{\rm d}49^{\rm m}13^{\rm s}\hspace{-5pt}.\hspace{2pt}0$ &  0.083 &$-0.035$& BF CAM   & M:  &\\
04265$+$5718 \dotfill & $04^{\rm h}26^{\rm m}31^{\rm s}\hspace{-5pt}.\hspace{2pt}9$ & $+57^{\rm d}18^{\rm m}12^{\rm s}\hspace{-5pt}.\hspace{2pt}7$ &  0.044 &$-0.073$& RV CAM   & SRb &\\
04554$+$4437 \dotfill & $04^{\rm h}55^{\rm m}29^{\rm s}\hspace{-5pt}.\hspace{2pt}8$ & $+44^{\rm d}37^{\rm m}07^{\rm s}\hspace{-5pt}.\hspace{2pt}0$ &  0.063 &$-0.051$&          &     &\\
05026$+$4447 \dotfill & $05^{\rm h}02^{\rm m}38^{\rm s}\hspace{-5pt}.\hspace{2pt}5$ & $+44^{\rm d}47^{\rm m}40^{\rm s}\hspace{-5pt}.\hspace{2pt}0$ &  0.065 &$-0.037$&          &     &\\
05176$+$3502 \dotfill & $05^{\rm h}17^{\rm m}35^{\rm s}\hspace{-5pt}.\hspace{2pt}0$ & $+35^{\rm d}02^{\rm m}24^{\rm s}\hspace{-5pt}.\hspace{2pt}0$ &  0.073 &$-0.007$& EE AUR   & Lb  &\\
06153$-$3100 \dotfill & $06^{\rm h}15^{\rm m}23^{\rm s}\hspace{-5pt}.\hspace{2pt}9$ & $-31^{\rm d}00^{\rm m}14^{\rm s}\hspace{-5pt}.\hspace{2pt}6$ &  0.083 &$-0.063$& EH CMA   & M   &\\
07186$-$1017 \dotfill & $07^{\rm h}18^{\rm m}36^{\rm s}\hspace{-5pt}.\hspace{2pt}2$ & $-10^{\rm d}17^{\rm m}03^{\rm s}\hspace{-5pt}.\hspace{2pt}0$ &  0.128 &$-0.047$& V 632MON & SR: &\\
07393$-$0403 \dotfill & $07^{\rm h}39^{\rm m}18^{\rm s}\hspace{-5pt}.\hspace{2pt}5$ & $-04^{\rm d}03^{\rm m}30^{\rm s}\hspace{-5pt}.\hspace{2pt}0$ &  0.036 &$-0.056$&          &     &\\
08196$+$1509 \dotfill & $08^{\rm h}19^{\rm m}36^{\rm s}\hspace{-5pt}.\hspace{2pt}9$ & $+15^{\rm d}09^{\rm m}11^{\rm s}\hspace{-5pt}.\hspace{2pt}1$ &  0.041 &$-0.076$& Z CNC    & SRb &\\
11538$+$5808 \dotfill & $11^{\rm h}53^{\rm m}54^{\rm s}\hspace{-5pt}.\hspace{2pt}0$ & $+58^{\rm d}09^{\rm m}00^{\rm s}\hspace{-5pt}.\hspace{2pt}0$ &  0.058 &$-0.065$& Z UMA    & SRb &\\
16418$+$5459 \dotfill & $16^{\rm h}41^{\rm m}52^{\rm s}\hspace{-5pt}.\hspace{2pt}0$ & $+54^{\rm d}59^{\rm m}48^{\rm s}\hspace{-5pt}.\hspace{2pt}0$ &  0.051 &$-0.055$& S DRA    & SRb &\\
16473$+$5753 \dotfill & $16^{\rm h}47^{\rm m}24^{\rm s}\hspace{-5pt}.\hspace{2pt}0$ & $+57^{\rm d}54^{\rm m}00^{\rm s}\hspace{-5pt}.\hspace{2pt}0$ &  0.091 &$-0.085$& AH DRA   & SRb &\\
17359$+$4555 \dotfill & $17^{\rm h}35^{\rm m}56^{\rm s}\hspace{-5pt}.\hspace{2pt}3$ & $+45^{\rm d}55^{\rm m}58^{\rm s}\hspace{-5pt}.\hspace{2pt}0$ &  0.044 &$-0.072$&          &     &\\
17473$+$4542 \dotfill & $17^{\rm h}47^{\rm m}22^{\rm s}\hspace{-5pt}.\hspace{2pt}0$ & $+45^{\rm d}42^{\rm m}54^{\rm s}\hspace{-5pt}.\hspace{2pt}0$ &  0.013 &$-0.081$& V 337HER & SRb &\\
18052$+$4326 \dotfill & $18^{\rm h}05^{\rm m}17^{\rm s}\hspace{-5pt}.\hspace{2pt}1$ & $+43^{\rm d}26^{\rm m}40^{\rm s}\hspace{-5pt}.\hspace{2pt}4$ &  0.001 &$-0.093$&          &     &\\
18064$+$4212 \dotfill & $18^{\rm h}06^{\rm m}26^{\rm s}\hspace{-5pt}.\hspace{2pt}0$ & $+42^{\rm d}12^{\rm m}54^{\rm s}\hspace{-5pt}.\hspace{2pt}0$ &  0.041 &$-0.067$& V529 HER & SR  &\\
18291$+$3836 \dotfill & $18^{\rm h}29^{\rm m}11^{\rm s}\hspace{-5pt}.\hspace{2pt}0$ & $+38^{\rm d}36^{\rm m}12^{\rm s}\hspace{-5pt}.\hspace{2pt}0$ &  0.017 &$-0.115$& KP LYR   & SR  &\\
18401$+$2854 \dotfill & $18^{\rm h}40^{\rm m}07^{\rm s}\hspace{-5pt}.\hspace{2pt}0$ & $+28^{\rm d}54^{\rm m}30^{\rm s}\hspace{-5pt}.\hspace{2pt}0$ &  0.076 &$-0.044$& FI LYR   & SRb &\\
18505$+$3327 \dotfill & $18^{\rm h}50^{\rm m}30^{\rm s}\hspace{-5pt}.\hspace{2pt}6$ & $+33^{\rm d}27^{\rm m}29^{\rm s}\hspace{-5pt}.\hspace{2pt}0$ &  0.042 &$-0.063$& HM LYR   & Lb  &\\
18512$+$3034 \dotfill & $18^{\rm h}51^{\rm m}12^{\rm s}\hspace{-5pt}.\hspace{2pt}8$ & $+30^{\rm d}34^{\rm m}08^{\rm s}\hspace{-5pt}.\hspace{2pt}0$ &  0.085 &$-0.043$&          &     &\\
19040$+$2416 \dotfill & $19^{\rm h}04^{\rm m}03^{\rm s}\hspace{-5pt}.\hspace{2pt}3$ & $+24^{\rm d}16^{\rm m}32^{\rm s}\hspace{-5pt}.\hspace{2pt}0$ &  0.015 &$-0.093$&          &     &\\
19194$+$1734 \dotfill & $19^{\rm h}19^{\rm m}28^{\rm s}\hspace{-5pt}.\hspace{2pt}2$ & $+17^{\rm d}34^{\rm m}14^{\rm s}\hspace{-5pt}.\hspace{2pt}0$ &  0.048 &$-0.045$& T SGE    & SRb &2)\\
19267$+$0345 \dotfill & $19^{\rm h}26^{\rm m}43^{\rm s}\hspace{-5pt}.\hspace{2pt}0$ & $+03^{\rm d}45^{\rm m}30^{\rm s}\hspace{-5pt}.\hspace{2pt}0$ &  0.051 &$-0.046$& V 858AQL & Lb  &\\
19306$+$0455 \dotfill & $19^{\rm h}30^{\rm m}39^{\rm s}\hspace{-5pt}.\hspace{2pt}0$ & $+04^{\rm d}55^{\rm m}12^{\rm s}\hspace{-5pt}.\hspace{2pt}0$ &$-0.012$&$-0.097$& V1293AQL & SRb &\\
19461$+$0334 \dotfill & $19^{\rm h}46^{\rm m}07^{\rm s}\hspace{-5pt}.\hspace{2pt}0$ & $+03^{\rm d}34^{\rm m}18^{\rm s}\hspace{-5pt}.\hspace{2pt}0$ &  0.022 &$-0.090$& WX AQL   & SRb &\\
20073$-$1041 \dotfill & $20^{\rm h}07^{\rm m}22^{\rm s}\hspace{-5pt}.\hspace{2pt}5$ & $-10^{\rm d}41^{\rm m}04^{\rm s}\hspace{-5pt}.\hspace{2pt}6$ &  0.046 &$-0.094$& $^\dagger$&    &\\
20094$-$1121 \dotfill & $20^{\rm h}09^{\rm m}29^{\rm s}\hspace{-5pt}.\hspace{2pt}3$ & $-11^{\rm d}21^{\rm m}21^{\rm s}\hspace{-5pt}.\hspace{2pt}1$ &  0.079 &$-0.050$&          &     &\\
20161$-$1600 \dotfill & $20^{\rm h}16^{\rm m}08^{\rm s}\hspace{-5pt}.\hspace{2pt}0$ & $-16^{\rm d}00^{\rm m}54^{\rm s}\hspace{-5pt}.\hspace{2pt}0$ &  0.085 &$-0.083$& AE CAP   & SR  &\\
20311$-$2325 \dotfill & $20^{\rm h}31^{\rm m}11^{\rm s}\hspace{-5pt}.\hspace{2pt}0$ & $-23^{\rm d}25^{\rm m}18^{\rm s}\hspace{-5pt}.\hspace{2pt}0$ &  0.024 &$-0.130$& AK CAP   & Lb  &\\
22073$+$7231 \dotfill & $22^{\rm h}07^{\rm m}23^{\rm s}\hspace{-5pt}.\hspace{2pt}0$ & $+72^{\rm d}31^{\rm m}24^{\rm s}\hspace{-5pt}.\hspace{2pt}0$ &$-0.008$&$-0.081$& DM CEP   & Lb  &\\

\hline
\end{tabular*}
Comment 2) means the same as table~1.
\vspace{6pt}\par\noindent
$\dagger$ SAO~163310 
\end{table}

%
%

\begin{table}[t]
\scriptsize

\begin{center}
Table~3. \hspace{4pt} 
Results of SiO maser observations for detected sources.
\end{center}

\vspace{6pt}
\tabcolsep 2pt
\begin{tabular*}{\columnwidth}{@{\hspace{\tabcolsep}
\extracolsep{\fill}}ll | rrrr | rrrr | rrrr}
\hline\hline\\[-6pt]

IRAS name & Date & \multicolumn{4}{l|}{$J=1-0, v=1$} & 
     \multicolumn{4}{l|}{$J=1-0, v=2$} &
     \multicolumn{4}{l}{$J=2-1, v=1$} \\
 &  &
$rms$ & $V_{\rm LSR}$ & $T_{\rm a}^{*}$ & $S_{\rm a}^{*}$ &
$rms$ & $V_{\rm LSR}$ & $T_{\rm a}^{*}$ & $S_{\rm a}^{*}$ &
$rms$ & $V_{\rm LSR}$ & $T_{\rm a}^{*}$ & $S_{\rm a}^{*}$ \\ \hline

02255$+$6903 & 5.20 & 0.04 & $ -29.7$ &  0.78 &   2.19 &   0.04 & $  -28.9$ &  0.92 &   3.16 &   0.07 & $ -31.1$ &  0.35 &  0.64 \\
04081$+$5832 & 5.20 & 0.02 & $-111.9$ &  0.15 &   0.47 &   0.02 & $ -112.9$ &  0.10 &   0.34 & $\cdot$& $  \cdot $ &$\cdot$&$\cdot$\\
06266$-$1148 & 6.11 & 0.02 & $  16.9$ &  0.23 &   0.55 &   0.02 & $   18.6$ &  0.09 &   0.19 &   0.05 & $  17.7$ &  0.65 &  1.16 \\
07019$-$1631 & 6.12 & 0.04 & $  53.7$ &  1.10 &   3.24 &   0.03 & $   55.4$ &  1.76 &   6.33 &   0.06 & $  54.0$ &  0.25 &  0.41 \\
07232$-$0544 & 5.20 & 0.04 & $  43.5$ &  3.59 &   6.29 &   0.04 & $   42.1$ &  3.05 &   6.13 &   0.07 & $  41.7$ &  0.79 &  1.58 \\
07268$-$0410 & 5.20 & 0.04 & $ 103.6$ &  0.43 &   0.74 &   0.04 & $  105.6$ &  0.22 &   0.75 & $\cdot$& $  \cdot $ &$\cdot$&$\cdot$\\
08186$+$1409 & 6.11 & 0.03 & $ -36.0$ &  0.31 &   0.10 &   0.03 & $  -38.3$ &  0.23 &   0.90 &   0.04 & $  \cdot $ &$\cdot$&$\cdot$\\
12341$+$5945 & 5.21 & 0.02 & $ -83.3$ &  0.10 &   0.38 &   0.02 & $  -89.0$ &  0.17 &   0.33 & $\cdot$& $  \cdot $ &$\cdot$&$\cdot$\\
18076$+$3445 & 5.23 & 0.04 & $   4.1$ &  1.40 &   3.67 &   0.04 & $    2.7$ &  1.03 &   4.41 &   0.06 & $   3.2$ &  0.26 &  0.65 \\
18156$+$0655 & 6.10 & 0.16 & $  37.4$ &  0.97 &   1.85 &   0.15 & $   37.3$ &  1.20 &   2.00 & $\cdot$& $  \cdot $ &$\cdot$&$\cdot$\\

18186$+$3143(a)& 5.22 & 0.03 & $  \cdot $ &$\cdot$& $\cdot$&   0.03 & $    \cdot$  &$\cdot$&$ \cdot$&   0.05 & $  19.1$ &  0.37 &  1.97 \\
18186$+$3143(b)& 5.23 & 0.03 & $  \cdot $ &$\cdot$& $\cdot$&   0.03 & $    8.0$ &  0.11 &   0.32 &   0.04 & $   4.7$ &  0.35 &  1.76 \\
18186$+$3143(c)& 6.11 & 0.03 & $  \cdot $ &$\cdot$& $\cdot$&   0.03 & $    5.6$ &  0.14 &   0.61 &   0.05 & $   8.4$ &  0.24 &$\cdot$\\
18222$+$3933 & 5.23 & 0.04 & $  18.9$ &  7.95 &  20.04 &   0.04 & $   20.6$ &  6.07 &  31.42 &   0.06 & $  17.9$ &  4.04 & 14.15 \\
18347$+$2600 & 5.24 & 0.03 & $  55.7$ &  0.24 &   0.85 &   0.03 & $   57.3$ &  0.19 &   0.58 &   0.05 & $  \cdot $ &$\cdot$&$\cdot$\\
18394$+$2845 & 5.24 & 0.04 & $  \cdot $ &$\cdot$& $\cdot$&   0.04 & $   -7.5$ &  0.20 &   0.65 & $\cdot$& $  \cdot $ &$\cdot$&$\cdot$\\
18561$+$1642 & 5.25 & 0.04 & $  73.9$ &  0.34 &   0.94 &   0.04 & $   76.4$ &  0.27 &   1.22 & $\cdot$& $  \cdot $ &$\cdot$&$\cdot$\\
19090$+$1746 & 6.10 & 0.10 & $  46.0$ &  0.79 &   1.40 &   0.10 & $   46.2$ &  1.25 &   3.66 & $\cdot$& $  \cdot $ &$\cdot$&$\cdot$\\
19158$+$1955 & 6.11 & 0.03 & $  49.0$ &  0.14 &   0.61 &   0.03 & $   49.4$ &  0.27 &   0.75 &   0.06 & $  \cdot $ &$\cdot$&$\cdot$\\
19172$+$1706 & 5.23 & 0.04 & $ -54.7$ &  0.32 &   0.91 &   0.03 & $  -54.2$ &  0.20 &   0.38 &   0.06 & $ -54.5$ &  0.39 &  1.40 \\
19285$+$4853 & 5.24 & 0.05 & $  12.0$ &  0.34 &   0.88 &   0.04 & $   11.2$ &  0.13 &   0.39 &   0.17 & $  \cdot $ &$\cdot$&$\cdot$\\
19308$+$0609 & 5.24 & 0.04 & $  46.6$ &  0.17 &   0.17 &   0.03 & $   \cdot$  &$\cdot$& $\cdot$&   0.14 & $  \cdot $ &$\cdot$&$\cdot$\\
19412$+$0337 & 5.23 & 0.08 & $ -31.1$ &  2.76 &  11.06 &   0.07 & $  -31.5$ &  2.01 &   8.01 &   0.12 & $ -30.8$ &  2.42 &  9.97 \\
19550$-$0201 & 5.23 & 0.09 & $  27.7$ & 89.96 & 180.57 &   0.08 & $   28.8$ & 43.45 & 153.45 &   0.13 & $  30.7$ &  4.99 & 16.29 \\
20077$-$0625 & 5.23 & 0.05 & $ -18.2$ &  0.42 &   0.93 &   0.05 & $  -18.2$ &  0.72 &   3.34 &   0.07 & $  \cdot $ &$\cdot$&$\cdot$\\
20215$+$6243 & 5.20 & 0.18 & $  17.2$ &  0.88 &   0.81 &   0.07 & $   18.9$ &  0.42 &   1.66 &   0.04 & $  \cdot $ &$\cdot$&$\cdot$\\
20234$-$1357 & 6.11 & 0.05 & $ -33.3$ &  1.90 &   6.78 &   0.05 & $  -33.1$ &  0.88 &   3.89 &   0.10 & $ -32.1$ &  0.98 &  1.90 \\
20296$-$2151 & 6.10 & 0.04 & $   6.4$ &  0.66 &   0.73 &   0.03 & $   \cdot $ &$\cdot$& $\cdot$& $\cdot$& $  \cdot $ &$\cdot$&$\cdot$\\
20305$+$6246 & 6.11 & 0.04 & $ -14.9$ &  0.85 &   1.66 &   0.03 & $  -14.5$ &  1.19 &   3.55 &   0.07 & $ -15.1$ &  0.22 &  0.37 \\

\hline
\end{tabular*}
Date is observed day in M.DD.
$Rms$ is root mean square of the antenna temperature (K), 
$V_{\rm LSR}$ is velocity at intensity peak (km~s$^{-1}$), 
$T_{\rm a}^{*}$ is antenna temperature at peak (K),
$S_{\rm a}^{*}$ is integrated intensity (K~km~s$^{-1}$).
\end{table}

%
%

\begin{table}[t]
\scriptsize
\begin{center}
Table~4.a. \hspace{4pt} 
The results of SiO maser observations for non-detected sources.
\end{center}
\vspace{6pt}
\tabcolsep 3pt
\begin{tabular*}{\columnwidth}{@{\hspace{\tabcolsep}
\extracolsep{\fill}}l rrrr}
\hline\hline\\[-6pt]

IRAS name & Date & $J=1-0, v=1$  &  $J=1-0, v=2$  &  $J=2-1, v=1$ \\
& & $rms$ & $rms$ & $rms$ \\[4pt]\hline\\[-6pt]

01010$+$7434 & 5.20 & 0.03 & 0.03 & 0.05 \\
01584$+$7103 & 6.11 & 0.03 & 0.03 & 0.07 \\
03478$+$6349 & 6.11 & 0.03 & 0.03 & 0.05 \\
04265$+$5718 & 5.20 & 0.03 & 0.03 & 0.04 \\
04265$+$5718 & 6.12 & 0.03 & 0.03 & 0.05 \\
04554$+$4437 & 6.12 & 0.03 & 0.03 & 0.04 \\
05026$+$4447 & 5.20 & 0.03 & 0.03 & 0.05 \\
05026$+$4447 & 6.12 & 0.02 & 0.02 & 0.04 \\
05176$+$3502 & 5.20 & 0.04 & 0.04 & 0.08 \\
06153$-$3100 & 6.12 & 0.04 & 0.04 & 0.07 \\
07186$-$1017 & 6.12 & 0.03 & 0.03 & 0.04 \\
07186$-$1017 & 6.11 & 0.03 & 0.03 & 0.05 \\
07393$-$0403 & 5.20 & 0.03 & 0.03 & 0.06 \\
08196$+$1509 & 6.12 & 0.02 & 0.02 & 0.04 \\
08196$+$1509 & 5.21 & 0.03 & 0.03 & 0.05 \\
11538$+$5808 & 5.21 & 0.03 & 0.03 & 0.05 \\
16418$+$5459 & 5.23 & 0.03 & 0.03 & 0.05 \\
16473$+$5753 & 5.23 & 0.03 & 0.03 & 0.05 \\
17359$+$4555 & 5.24 & 0.05 & 0.05 & $\cdot$ \\
17473$+$4542 & 5.23 & 0.03 & 0.03 & 0.05 \\
18052$+$4326 & 5.22 & 0.03 & 0.03 & 0.05 \\
18064$+$4212 & 5.23 & 0.04 & 0.04 & 0.07 \\
18064$+$4212 & 5.22 & 0.04 & 0.04 & 0.07 \\
18291$+$3836 & 5.23 & 0.03 & 0.03 & 0.06 \\
18291$+$3836 & 6.11 & 0.06 & 0.07 & $\cdot$ \\
18401$+$2854 & 6.11 & 0.04 & 0.04 & 0.08 \\
18401$+$2854 & 5.23 & 0.03 & 0.03 & 0.07 \\
18505$+$3327 & 6.11 & 0.04 & 0.04 & 0.09 \\
18512$+$3034 & 5.25 & 0.07 & 0.07 & $\cdot$ \\
19040$+$2416 & 5.23 & 0.04 & 0.03 & 0.06 \\
19194$+$1734 & 6.10 & 0.05 & 0.05 & $\cdot$ \\
19267$+$0345 & 5.24 & 0.03 & 0.03 & 0.06 \\
19267$+$0345 & 6.10 & 0.03 & 0.03 & $\cdot$ \\

\hline
\end{tabular*}
$rms$ (root mean square) is in the unit of K.
\end{table}

%
%
\begin{table}[t]
\scriptsize
\begin{center}
Table~4.b. \hspace{4pt} {\it Continued.}
\end{center}
\vspace{6pt}
\tabcolsep 3pt
\begin{tabular*}{\columnwidth}{@{\hspace{\tabcolsep}
\extracolsep{\fill}}l rrrr}
\hline\hline\\[-6pt]

IRAS name & Date & $J=1-0, v=2$  &  $J=1-0, v=1$  &  $J=2-1, v=1$ \\
 & & $rms$ & $rms$ & $rms$ \\[4pt]\hline\\[-6pt]

19306$+$0455 & 5.24 & 0.03 & 0.03 & 0.06 \\
19461$+$0334 & 5.24 & 0.03 & 0.03 & 0.09 \\
19461$+$0334 & 5.23 & 0.04 & 0.04 & 0.06 \\
20073$-$1041 & 5.23 & 0.04 & 0.03 & 0.06 \\
20094$-$1121 & 5.24 & 0.03 & 0.03 & 0.07 \\
20161$-$1600 & 5.24 & 0.03 & 0.03 & 0.06 \\
20161$-$1600 & 6.11 & 0.03 & 0.03 & 0.06 \\
20311$-$2325 & 5.23 & 0.03 & 0.02 & 0.05 \\
22073$+$7231 & 5.20 & 0.03 & 0.03 & 0.05 \\
22073$+$7231 & 6.11 & 0.03 & 0.03 & 0.08 \\

\hline
\end{tabular*}
\end{table}

%
%

\begin{table}[t]
\begin{center}
Table~5. \hspace{4pt} 
The distribution of detected source and non-detected source
at each variable type.
\end{center}
\vspace{6pt}
\tabcolsep 3pt
\begin{tabular}{@{\hspace{\tabcolsep}
\extracolsep{\fill}}lrr}
\hline\hline\\[-6pt]
Type & Detected & Non-detected \\[4pt]\hline\\[-6pt]

Mira    \dotfill & 15 & 2 \\ 
SR      \dotfill &  2 & 14\\
Lb      \dotfill &  1 & 5  \\
Unknown \dotfill & 9 & 11  \\ \hline
Total & 27 & 32 \\ 
\hline
\end{tabular}
\end{table}

%
%

\begin{table}[t]
\tiny
\begin{center}
Table~6. 
The observations of near-infrared water spectra
and SiO masers.
\end{center}
\vspace{6pt}
\begin{tabular*}{\columnwidth}{@{\hspace{\tabcolsep}
\tabcolsep=0pt
\extracolsep{\fill}}lll| ll |lll|l|l }
\hline\hline\\[-6pt]

Name      &  & Type & $N$(H$_2$O) cm$^{-2}$ & $T_{\rm ex}$(H$_2$O) K 
& \multicolumn{3}{c|}{SiO Masers}  & $d$ (pc) & ref.\\ 
          &  &       &                      &
           & $J$=1-0 & $J$=1-0 & $J$=2-1 &  &\\
& & & & &  $v=1$   & $v=2$   & $v=1$   &  &\\ \hline

AK~Cap      & $\dotfill$ & Lb     & $5\times 10^{19}$ & 1000--1500    & N & N & N & 497 & Paper~I, this work\\ 
SAO~163310  & $\dotfill$ & $\cdot$& $1\times 10^{20}$ & 1000--1500    & N & N & N &     & Paper~I, this work \\
V~Hor       & $\dotfill$ & SRb    & $1\times 10^{20}$ & 1000--1500    & N &    &  & 336 & Paper~I, (4)\\ \hline

$\beta$~Peg & $\dotfill$ & Lb     & $7\times 10^{18}$ & 1250          & N & N &   &  61 & (1), (5)\\
g~Her       & $\dotfill$ & SRb    & $2\times 10^{19}$ & 1250          & N & N &   & 111 & (1), (5)\\
SW~Vir      & $\dotfill$ & SRb    & $3\times 10^{19}$ & 1250          & N & N & D & 143 & (1), (5), (6)\\
R~Leo & warm           & Mira    & $<5\times10^{19}$--$3\times10^{20}$ & 1700          & D &   & D & 101 & (2), (7), (8)\\
      & cool           &         & $2\times10^{21}$--$2\times10^{22}$ & 1150 -- 1200 & & & & &\\ 
$o$~Cet & hot          & Mira	 & $3\times10^{21}$   & 2000          & D &   & D & 128 & (3), (8), (9)\\
        & cool         & 	 & $3\times10^{20}$   & 1400          & & & & \\
Z~Cas   & hot          & Mira	 & $3\times10^{21}$   & 2000          & & & & & (3), (10) \\
        & cool         & 	 & $1\times10^{21}$   & 1200          & & & & \\\hline

\end{tabular*}

`$d$' indicates the distance to the star
from the Hipparcos and Tycho catalogues (1997).
Hinkle \& Barnes (1979) and Yamamura et al. (1999b)
measured water spectra with
two components of layers, and both layers are indicated individually.
The phase dependence of density and temperature
is detected in R~Leo.
In the column in SiO maser, D is detection and N is non-detection. 
References (1)--(3) are for the water observations
and reference (4)--(10) are for the SiO maser observations.

\vspace{6pt}\par\noindent
(1) Tsuji et al. (1997)
\vspace{6pt}\par\noindent
(2) Hinkle \& Barnes (1979)
\vspace{6pt}\par\noindent
(3) Yamamura et al. (1999b)

\vspace{6pt}\par\noindent
(4) Allen et al. (1989)
\vspace{6pt}\par\noindent
(5) Alcolea et al. (1990)
\vspace{6pt}\par\noindent
(6) Herpin et al. (1998)
\vspace{6pt}\par\noindent
(7) Heske (1989)
\vspace{6pt}\par\noindent
(8) Nyman \& Olofsson (1986)
\vspace{6pt}\par\noindent
(9) Buhl et al. (1974)
\vspace{6pt}\par\noindent
(10) Spencer et al. (1981)

\end{table}

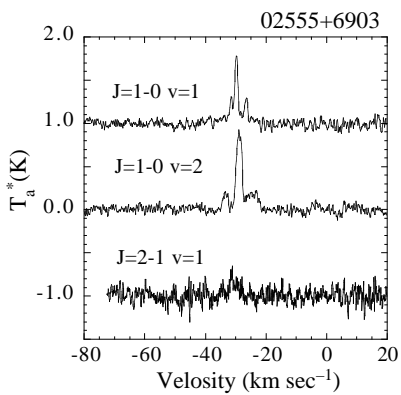
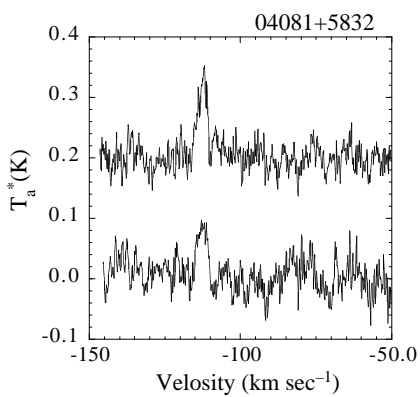
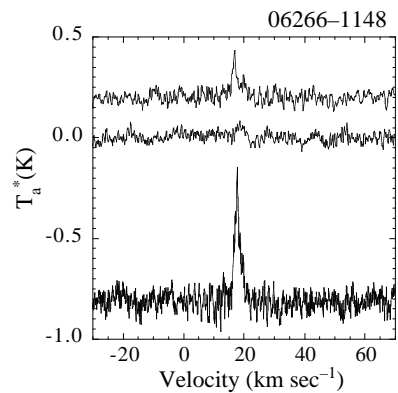
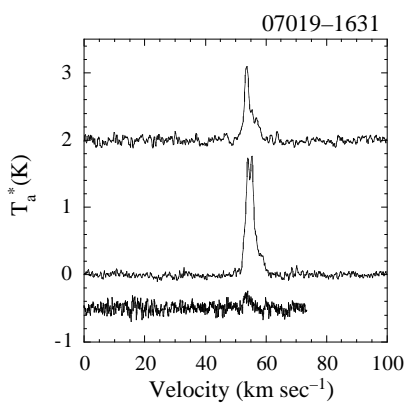
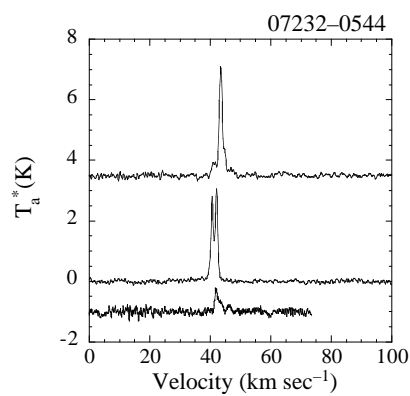
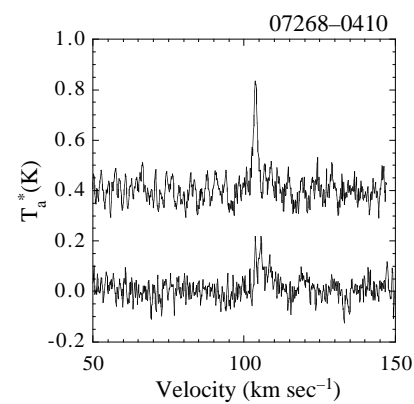
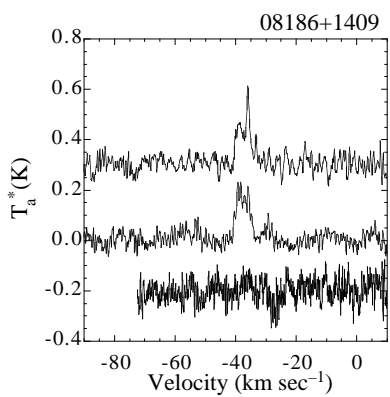
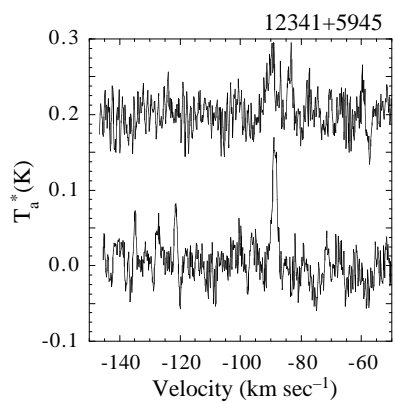
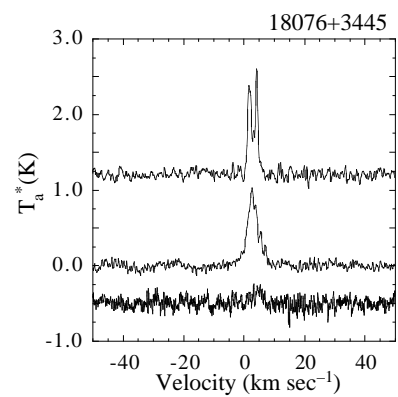
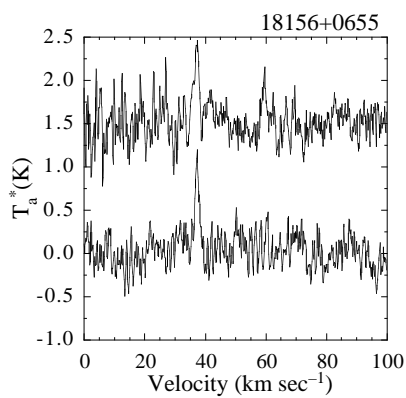
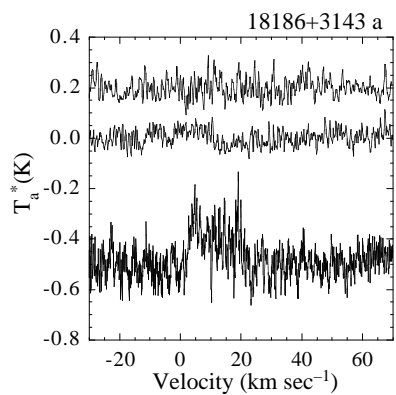
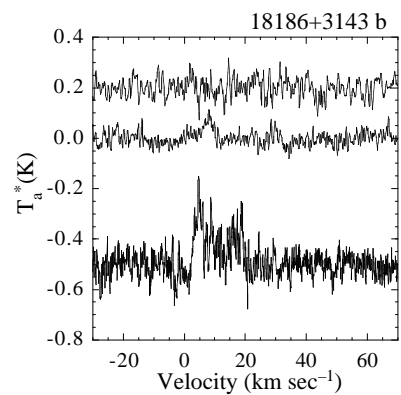
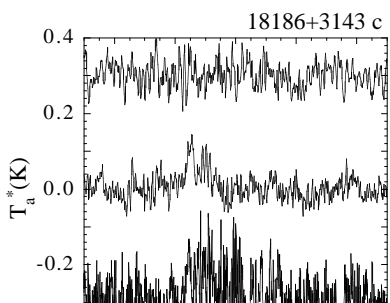
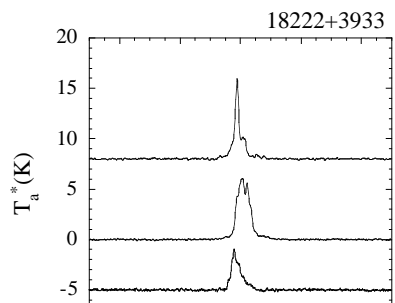
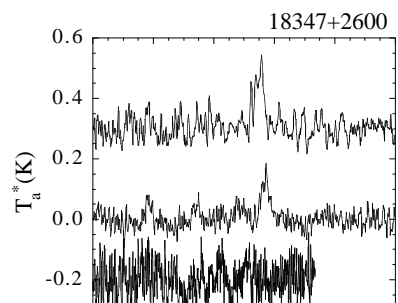
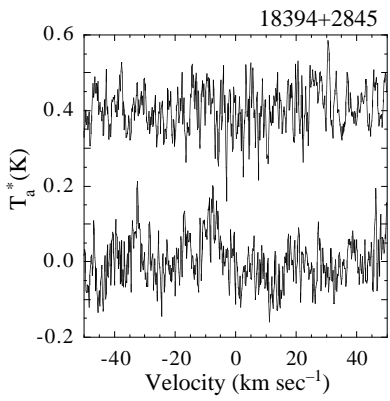
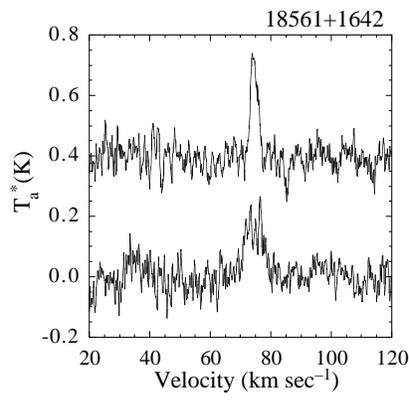
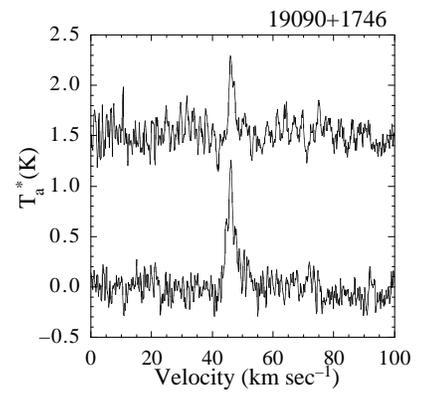
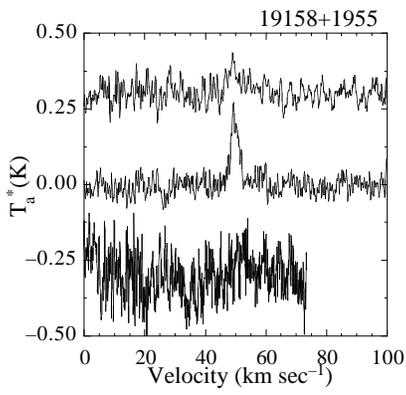
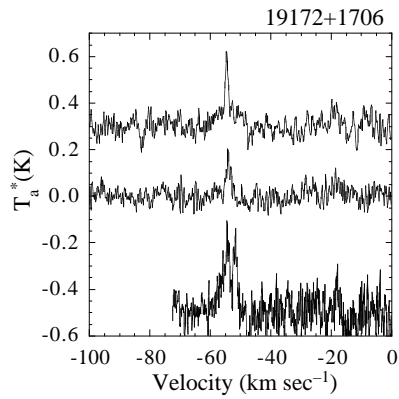
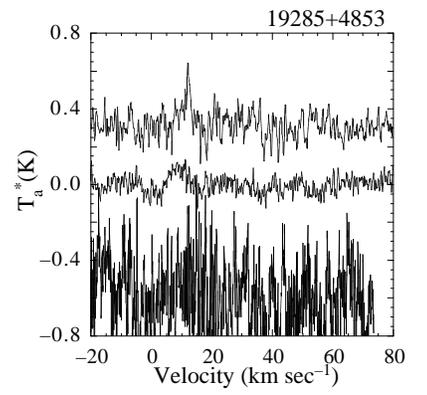
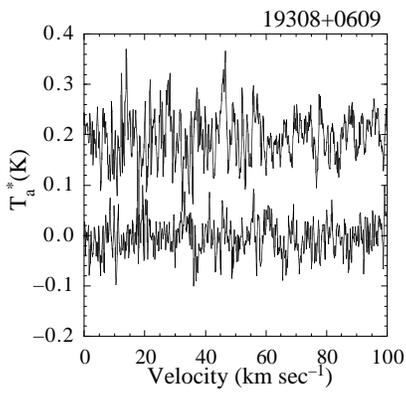
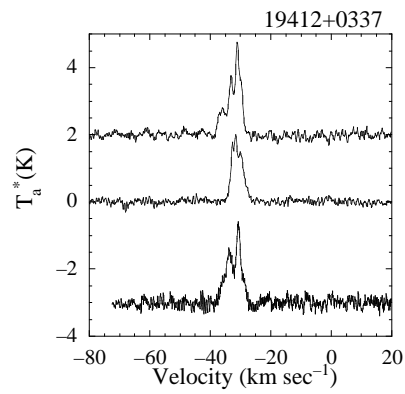
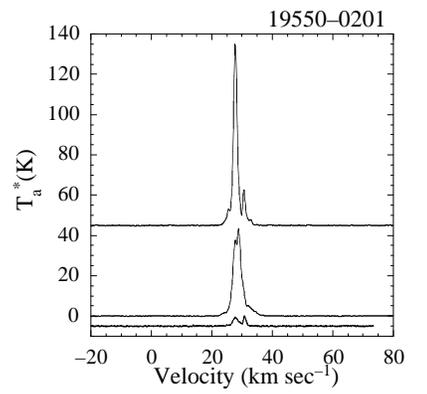
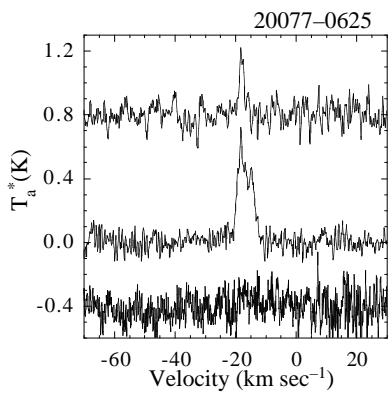
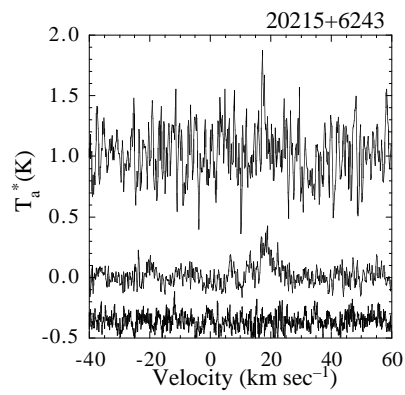
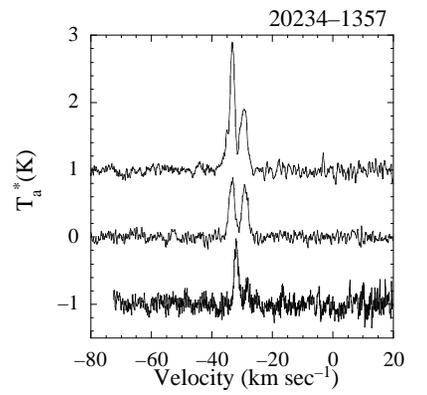
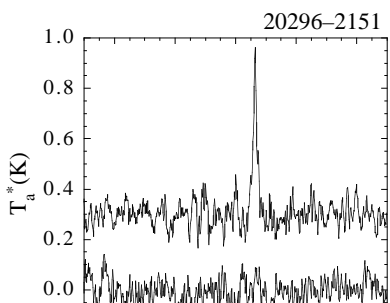
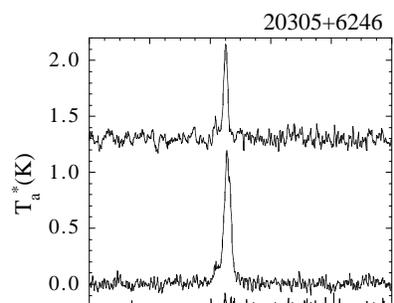
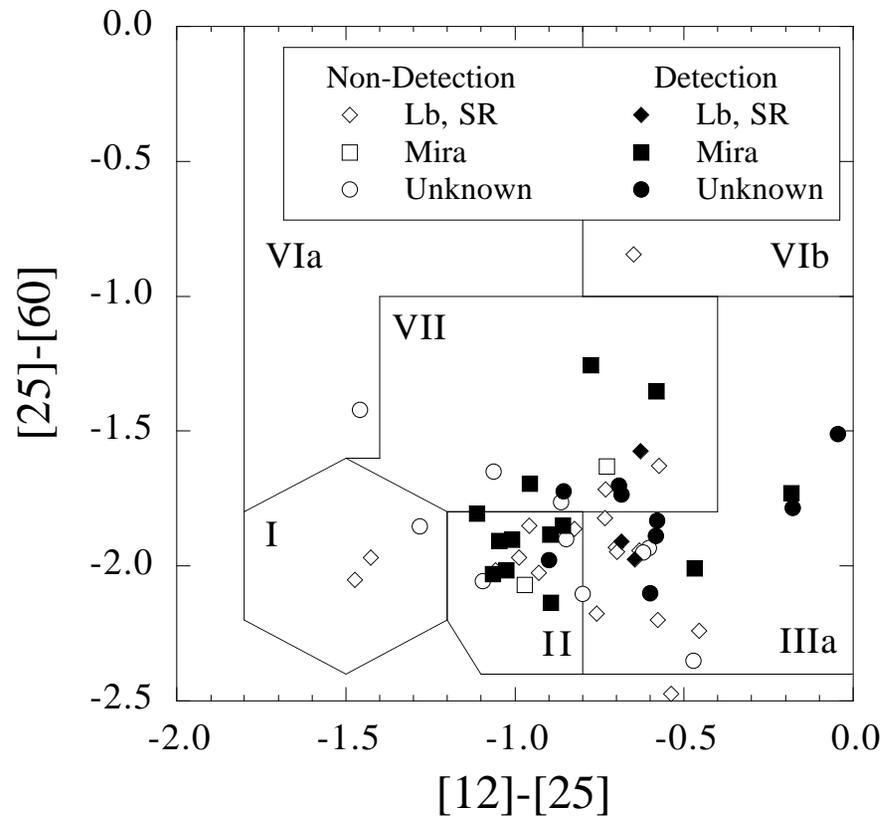
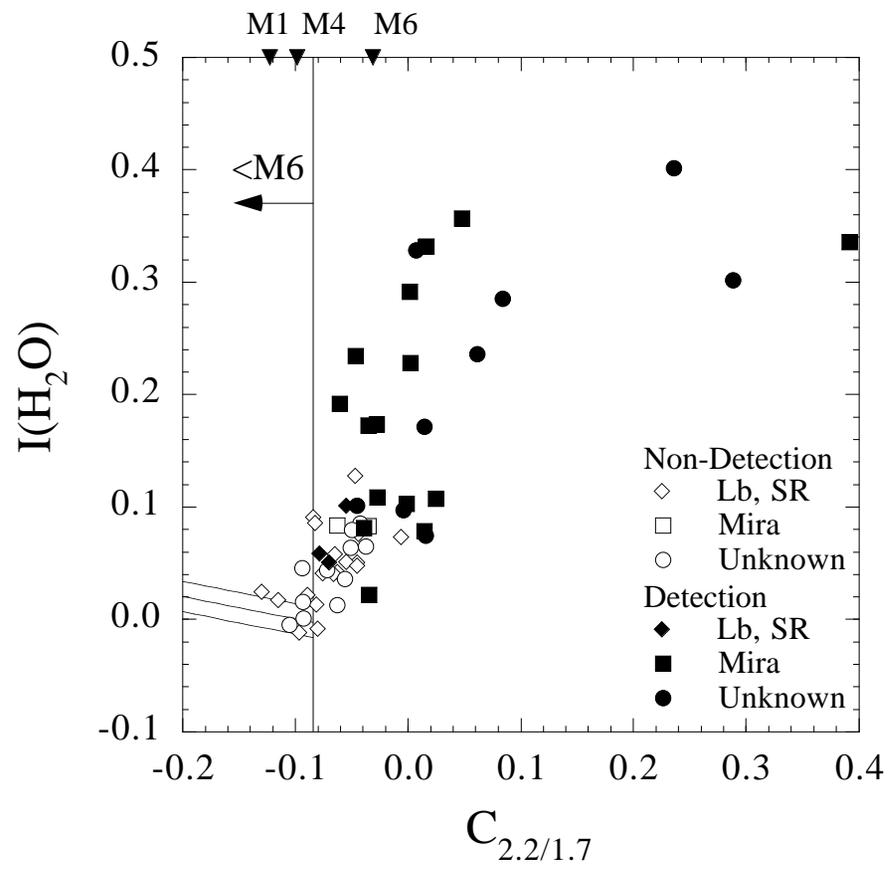
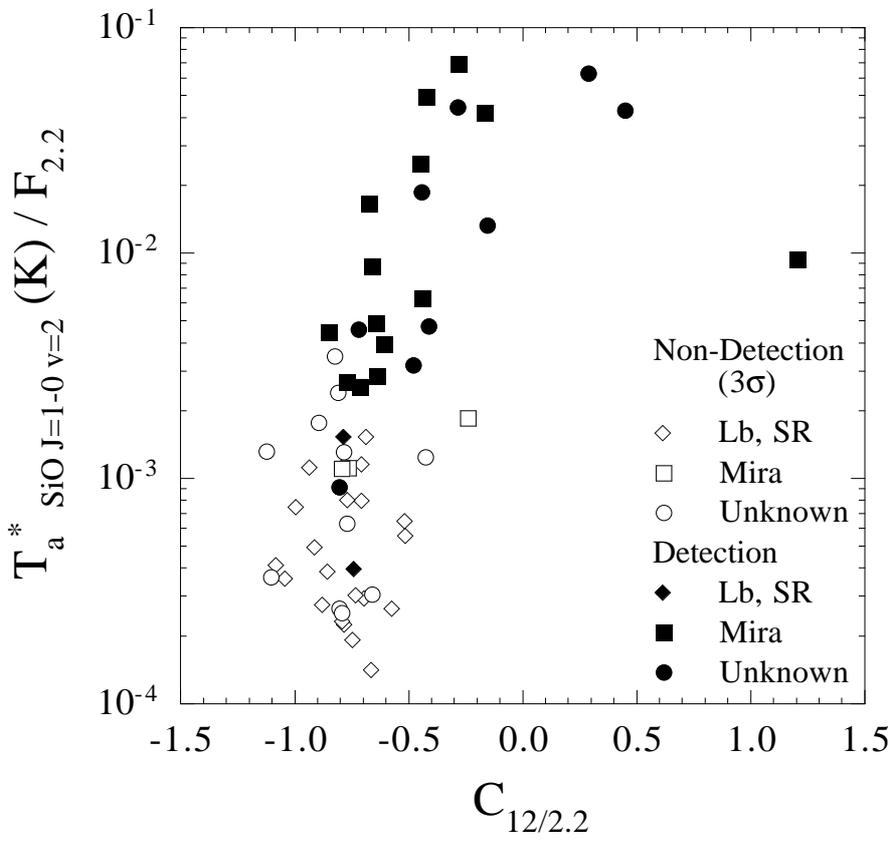
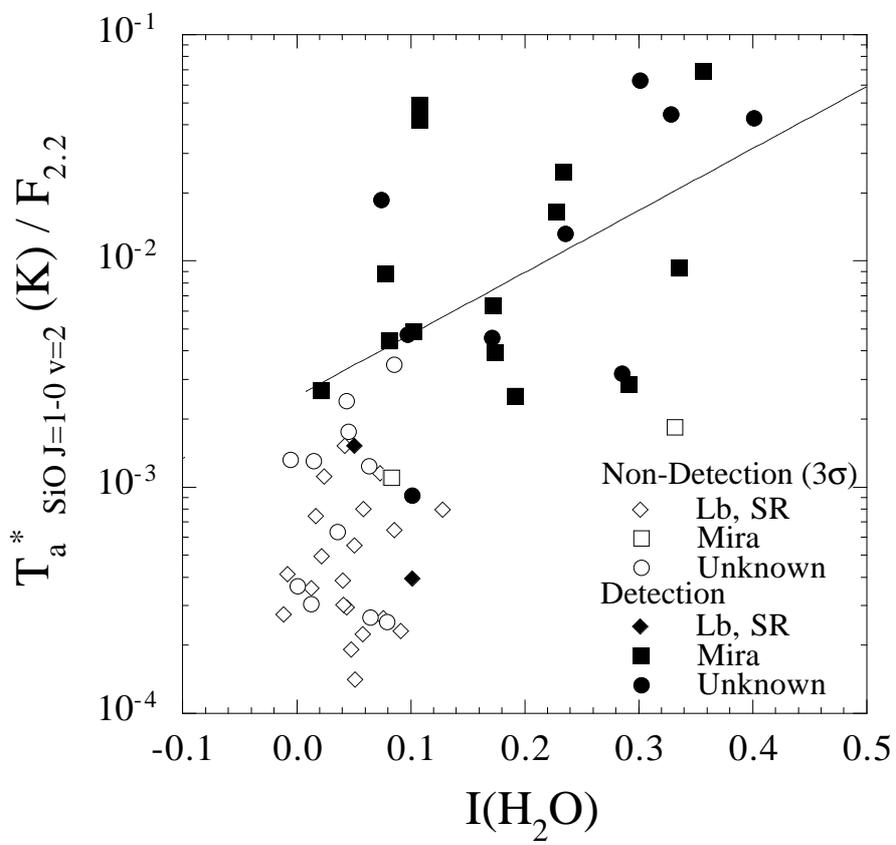
\end{document}